\documentclass[runningheads]{llncs}
\usepackage[T1]{fontenc}
\usepackage{graphicx}
\usepackage{array}
\usepackage{cite}
\usepackage{chngcntr}
\usepackage{appendix}
\begin{document}
%

% old title: AI and avatars: The effect of appearance and disclosure on user motivation
\title{The Impacts of AI Avatar Appearance\\
and Disclosure on User Motivation}
\titlerunning{AI and Avatars}
\author{Boele Visser\inst{1}\orcidID{0009-0002-7504-6495} \and
Peter van der Putten\inst{1}\orcidID{0000-0002-6507-6896} \and 
Amirhossein Zohrehvand\inst{1}\orcidID{0000-0002-7473-6702}}

\authorrunning{B. Visser et al.}
\institute{Leiden Institute of Advanced Computer Science, Leiden University\\ P.O. Box 9512, 2300 RA Leiden, The Netherlands}

\maketitle              % typeset the header of the contribution
\begin{abstract}
  This study examines the influence of perceived AI features on user motivation in virtual interactions. AI avatars, being disclosed as being an AI, or embodying specific genders, could be used in user-AI interactions. Leveraging insights from AI and avatar research, we explore how AI disclosure and gender affect user motivation. We conducted a game-based experiment involving over 72,500 participants who solved search problems alone or with an AI companion. Different groups experienced varying AI appearances and disclosures. We measured play intensity. Results revealed that the presence of another avatar led to less intense play compared to solo play. Disclosure of the avatar as AI heightened effort intensity compared to non-disclosed AI companions. Additionally, a masculine AI appearance reduced effort intensity.

\end{abstract}

\section{Introduction}

With the rise of digital worlds~\cite{Stefko, veltri2014gender, verhagen2012understanding, zohrehvand2022fifty}, humans are increasingly interacting with different forms of Artificial Intelligence (AI) in digital spaces, for example, in the form of the recommended system in social media, chatbots or avatars. Similar to the offline world, the perceived gender of the entity that we are interacting with shapes our behaviour and motivation~\cite{adam2021ai, mcdonnell2019chatbots, vanneste2024artificial}. Past research on motivation in offline settings argues that gender differences in motivation arise from the social and cultural expectations, experiences and roles that are associated with being male or female~\cite{meece2009gender}. Gender roles and stereotypes can shape an individual’s beliefs, values and behaviours, including their motivation~\cite{meece2009gender}. These predispositional features are almost immutable.

However, contrary to the offline world, the gender of the agent that we are interacting with, i.e. the AI agent, is customizable. The AI gender appearances could be chosen on a scale, allowing us to experiment with immutable characteristics and observe their impact on user behaviour. Building on social exchange theory~\cite{pillemer2018friends, westphal1997defections} and trust and agency literature~\cite{vanneste2024artificial} we argue that users playing together with an avatar exhibit a higher degree of persistence in comparison with playing alone. Further, we argue that users are more motivated to play with an agent if they attribute more agency to or perceive the gender of the agent to be of a more cooperative type, they exhibit higher degrees of persistence. Therefore, we suggest that users play less intensely when they are playing with an AI in comparison to a human. Likewise, we argue that the disclosure that an avatar is actually AI-driven reduces the intensity which with the user plays, just as what happens when AI systems in the physical world are disclosed ~\cite{tong2021janus, keppeler2024no}.

To research this, we performed an experiment in a game. In the game, published on Roblox  (a virtual world), players have to perform a simple task, which can be done together with others. Inside this game, over 72,500 participants played 15 levels of a simple search task. The participants either played alone or together with an AI avatar, from which the gender appearance was randomly chosen. For half of the participants playing together with an AI, it was not disclosed that they were playing with an AI. We measured how intense participants played the experiment by dividing the time before leaving the experiment by the number of levels completed. We used this measurement to compare the difference in motivation between different treatment groups, to see what effect the AI gender appearance has on motivation in this particular context. The findings of this study expand our understanding of social exchange theory in the online world~\cite{pillemer2018friends} and trust in AI~\cite{vanneste2024artificial}  by showing how the perceived gender of a non-human agent influences human motivation and trust.

\section{Background}
Motivation is defined as ``the degree and type of effort that an individual exhibits in a behavioural situation''~\cite{perry1982factors}. There are two types of effort: intrinsic motivation and extrinsic motivation~\cite{ryan2000intrinsic}. Intrinsic motivation occurs when an individual finds something interesting or enjoyable, while extrinsic motivation occurs when an individual is motivated by the outcome of a task. While intrinsic motivation is more about the process, extrinsic motivation is more about the outcome. Past research shows that different demographics show categorically different motivational patterns or the demographics of others might influence their motivation. According to a literature review~\cite{sekhar2013literature}, motivation drivers vary among age groups.

\subsection{Social interactions with AI}
Interactions with artificial intelligence (AI) are becoming increasingly common.  Considering that users can give agency to the AI agent, the characteristics of the AI can influence how people interact~\cite{kim2019effects, zohrehvand2020m} and trust AI based on their ability to think, plan and act~\cite{vanneste2024artificial}. Past research shows that even the features of how the AI agent is described to users can influence the interaction and the user~\cite{Mols2020-wo}. Within the interaction itself, the form factor (i.e. written, voice, physical) of the AI changes how participants conform to the AI~\cite{schreuter2021trust}. In these interactions, the perceived gender of the AI influences the compliance with the AI~\cite{lee2000can, mcdonnell2019chatbots}. To sum, past research shows that various factors influence how people interact with an AI. Increasingly, virtual worlds are where many of these interactions with AI could take place.

Virtual spaces are expanding beyond gaming communities to offer courses, stores, and financial services, allowing consumers to discuss their interactions with firms, and for many more aspects of everyday life to take place online~\cite{verhagen2012understanding, zohrehvand2024generalizing}.  Both intrinsic and extrinsic motivation, measured by factors such as escapism, visual attractiveness, perceived usefulness, and entertainment value, impact users' attitudes towards using virtual worlds~\cite{verhagen2012understanding}.

In virtual worlds, users are typically represented by avatars, which are graphical representations of themselves~\cite{holzwarth2006influence}. Users can design their avatars to represent their actual appearance or their desired identity in the virtual world~\cite{belisle2010avatars}.
The appearance of an avatar can influence the behaviour and motivation of the user, especially when the user can identify with their avatar~\cite{you2013feel, birk2016fostering, baylor2011design}. The appearance of an avatar can also influence how a user is perceived by others in virtual worlds~\cite{belisle2010avatars, shih2023you, van2015does}. Particularly, avatar gender can also influence how a user is perceived and behaves in virtual worlds~\cite{veltri2014gender}.

Users, similar to the offline world, can interact with others, represented by some avatar, in a virtual world. Past research suggests that cooperating with other players motivates players to use a virtual world: When an educational video game app added a feature where users could asynchronously cooperate, the app usage increased 217\%~\cite{featherstone2019unicraft}.
Synchronous cooperation also motivates players: players of video games are motivated to play an exergame again more when they play together (in cooperation) with another player in the same physical space than when they played alone~\cite{peng2013playing}. When the AI behaves like a player, or when the player cannot distinguish an AI agent from a human player, there should be no difference between the interaction with a player and with an AI, thus the effects of cooperating with another human could carry over to cooperating with an AI. Therefore, we suggest the following:

H1: Participants who play with an AI avatar (treatment group, regardless of disclosure) have a higher intensity of playing than those who play alone (control group).

\subsection{Relationship between AI avatar gender and motivation}
The gender of others who someone interacts with influences motivation~\cite{conti2001impact}. Some research claims that men are motivated by competition and they believe they outperform others~\cite{conti2001impact, thaler2021gender}, so men will want to keep ahead of others and are motivated to do so quickly. Other research claims that women behave more prosocially~\cite{soutschek2017dopaminergic}, so women might want to support others and continue to keep interacting longer, which will make them play longer.

Since an AI is perceived to be a social actor by humans~\cite{kim2019effects}, the characteristics of the AI determine how humans interact with the AI~\cite{kim2019effects} and the perceived gender of an AI having an influence on people’s motivation during textual interactions~\cite{mcdonnell2019chatbots}, the gender effects may carry on when interacting with an AI. We therefore suggest the following:

H2: Participants playing with an AI avatar that appears more masculine have a higher intensity of playing.

\subsection{Relationship between AI disclosure and motivation}
When interacting with an AI, the user could know whether the one they are interacting with is an AI or not (which implies it is a human). This AI disclosure affects how users perceive an AI system~\cite{tong2021janus, keppeler2024no}. Employees who get performance feedback created by an AI get a negative perception of this feedback when they know the feedback is created by an AI, which makes them less motivated by the feedback~\cite{tong2021janus}. Additionally, Job candidates are less interested in an offer when their application is reviewed by an AI system and therefore are less motivated~\cite{keppeler2024no}. Note that the framing that is used when introducing the AI may have an influence on trust and conformance \cite{Mols2020-wo}). This lets us believe that AI disclosure causes a lower motivation of users, and therefore we suggest the following:

H3: Intensity of playing is negatively associated with participants knowing that they are playing with an AI avatar, in comparison to playing without the AI being disclosed.

\section{Methods}
In this section, we will discuss the methodology of the experiment we performed. We preregistered this experiment at the Open Science Foundation.

\subsection{Organizational setting}
The setting for this experiment is a game on Roblox. Roblox is an online gaming platform that allows users to create, share and play their own virtual experiences. This platform has over 150 million monthly active users. The game had more than 50.000 active daily users at the time of the experiment. The idea behind the game is simple: players get placed in a level (a room), and their task is to find a hidden object inside that room. Once they find the object they click on it, and they progress to the next level. 

The player base of the game is very diverse. Players come from around the world, with different age groups, languages and devices. This makes it a good fit to research the impact of an avatar’s gender on motivation in a very diverse player base.

\subsection{Participants}
Participants are players of the Game X. Players could join the experiment by selecting a special mode called ‘Science Mode’. Only the data from plays during the first two weeks are used, although it is possible to join the ‘Science Mode’ after the experiment’s duration to be able to review the procedure. After participating in the experiment, players receive a cosmetic item and some in-game virtual currency.

Participants could participate in the experiment multiple times, but only their first participation will be taken into account for the analysis.

Due to the privacy regulations of Roblox, it is not possible to review the age and gender of the participant. However, data about age is available on a generic level of the game, where we could see the distribution of players in age groups. It is most likely that the participants follow the same demographics as the game itself. Players from the game come from around the world, which makes the participants very diverse and eliminates the WEIRD society bias from the data~\cite{henrich2010weirdest}.

\subsection{Procedure}
When participants join the experiment, they get an information screen that they are going to participate in an experiment. One-third of the participants are in the control group and will participate alone. The other two-thirds are in the subject group and will participate in an AI avatar. Within the subject group, one-half was told they were playing with an AI (so one-third of all participants), while the other half (also one-third) was not told they were playing with an AI.

After the explanation, participants select their avatar. Participants were completely free to select the avatar they wanted; the question asked was ‘Select your avatar’. The avatars differ in gender appearance. The avatars were always displayed in the same order, from left to right (See Figure~\ref{figA}). Participants play with their selected avatar during the experiment, and participants are asked to rate this avatar on a 5-point scale ranging from ‘very feminine’ to ‘very masculine’.
% \begin{figure}[tbp]
% \begin{minipage}{\linewidth}
%       \centering
%       \begin{minipage}{0.45\linewidth}
         
%         \includegraphics[width=\linewidth]{Figures/FigureA.png}
%         \caption{Avatar selection screen.} \label{figA}
%       \end{minipage}
%       \hspace{0.05\linewidth}
%       \begin{minipage}{0.45\linewidth}
%         \includegraphics[width=\linewidth]{Figures/FigureB.png}
%         \caption{Avatar rating screen.} \label{figB}
        
%       \end{minipage}
%   \end{minipage}
%   \end{figure}

\begin{figure}[tbp]
\centering
\includegraphics[width=0.8\textwidth]{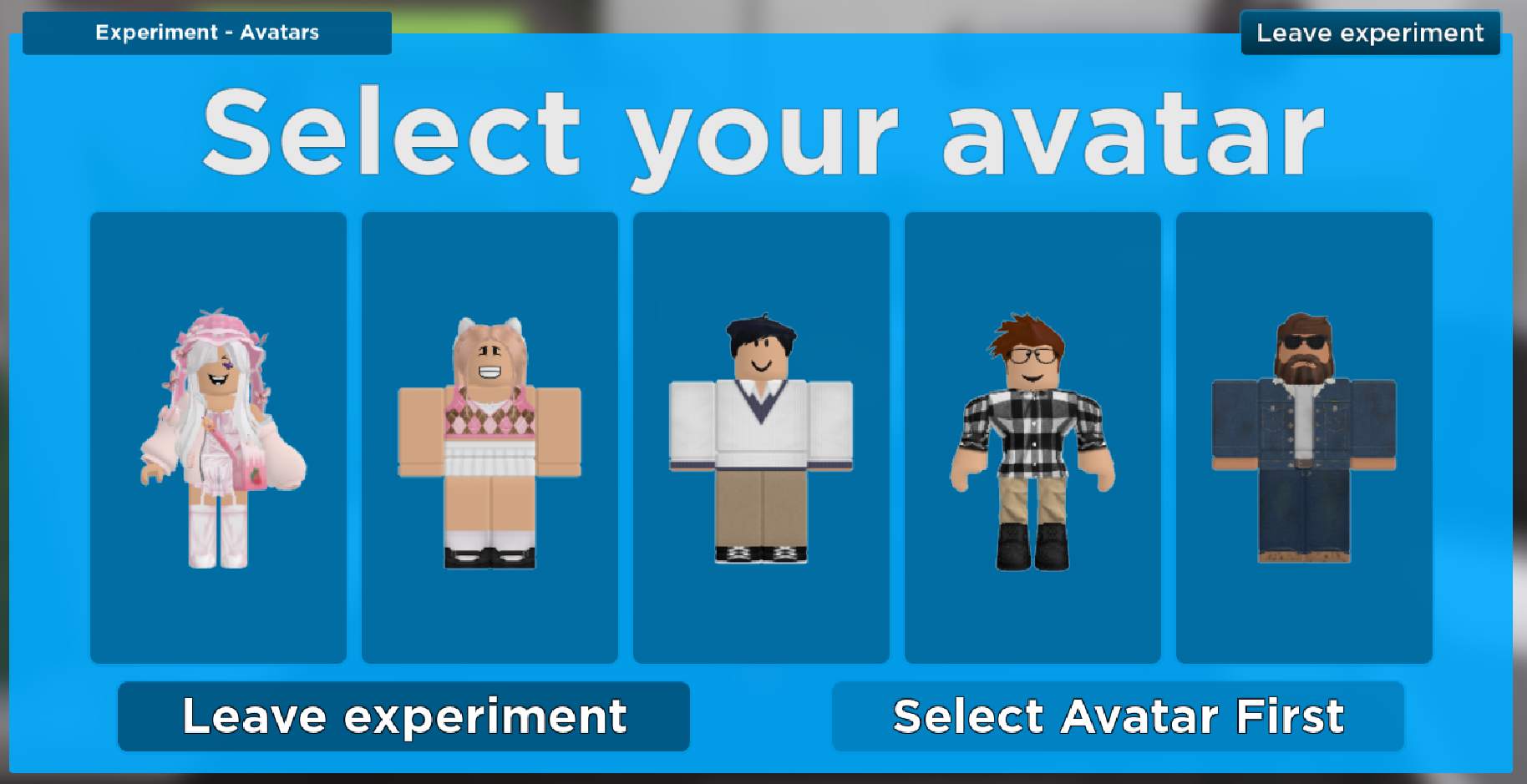}
\caption{Avatar rating screen.} \label{figB}
\end{figure}

After they select the avatar, they play the ‘Science Mode’, which consists of 15 stages. In this mode, players cannot see other participants, except the AI avatar controlled by the computer when participants are in the subject group. This AI avatar is also called a non-player character (NPC).

The appearance of the NPC is one of the 5 different avatars shown during the self-selection process. The appearance of the NPC is selected randomly out of the five preset avatars, regardless of the avatar chosen by the player. The NPC wanders around the room to preset positions and pretends to be searching for the button. When the participant finds the button, the NPC teleports to the next level within a few seconds. Therefore, the NPC is always playing at the same level as the participant, which gives the participant the feeling that the NPC is playing with them.

The difficulty and level design is the same for all participants, as is the behaviour of the NPC. The independent variables that are changed are playing alone vs. playing with an NPC, the disclosure of the NPC being an AI and the (gender) appearance of the NPC.

Before joining the experiment, participants are informed about its details. The experiment is non-intrusive and participation is voluntary, with the option to leave at any time without giving a reason. Participants are informed about the treatment type directly after participation. The AI is programmed to never harm participants, but rather to encourage and help them to complete the task while remaining friendly at all times. No personal identifier data was stored about the user (data collection has been fully anonymized).

\subsection{Measures}
We proxy motivation based on how intensely players engage with the experiment and how likely it is that they complete all the stages. We measure the seconds spent per level by calculating the time before leaving the experiment divided by the number of levels completed. A lower value of seconds spent per level therefore implies that the participant spends less time on a level on average, and therefore engages more intensely with the experiment. The motivation of the participants is measured by using a performance-based behavioural measure. Performance-based behavioural measures indicate both a higher outcome-focused motivation (extrinsic motivation) and process-focused motivation (intrinsic motivation)~\cite{toure2014measure}.

In addition, we use a dummy variable to indicate if a participant has completed all 15 stages. The time before ending the participation is also measured - participation can end by either completing the 15 stages or leaving earlier. The speed at which players complete the experiment therefore indicates if the avatar's appearance influences either the intrinsic process-focused motivation or extrinsic outcome-focused motivation.

To get a more reliable measure for the gender of the avatars, we asked each participant to rate their avatar on a 5-point scale, with 1 being ‘very feminine’  and 5 being ‘very masculine’. This gives an average score for each avatar between 1 and 5. To simplify the analysis, we calculated the average rating score of the avatar. Given that the average of our 1-5 scale is 3, we mapped the avatar to either feminine when the average score is below 3 or masculine when the average score is above 3. If a score is exactly 3, the avatar is treated as having no gender and is excluded from the analysis about gender.

\subsection{Analytical strategy}
To test the hypotheses, we used Ordinary Least Squares (OLS) from the statsmodels package~\cite{seabold2010statsmodels} in Python. For each variable, the dependent variable is the seconds spent per level, modelled using OLS. The independent variable is AI avatar presence, AI avatar masculinity and AI disclosure for respectively H1, H2 and H3.

\section{Results}
During the 2-week time window, 126,013 participants started the experiment. From this group, 8488 participants (6.74\%) completed all levels of the experiment.

\begin{figure}[tbp]
\centering
\includegraphics[width=0.6\textwidth]{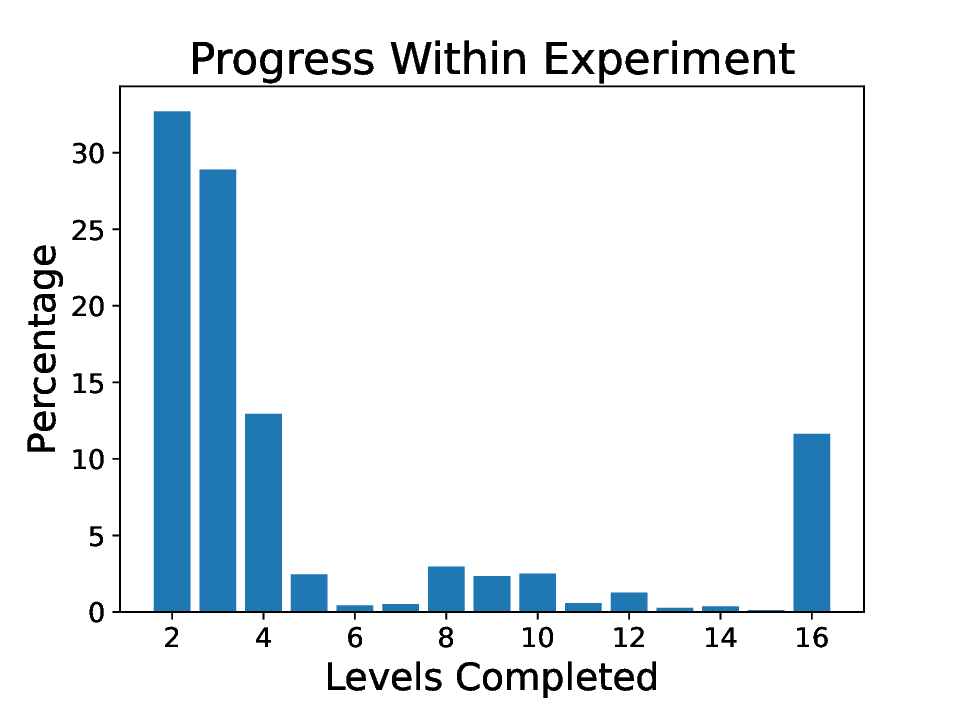}
\caption{Progress of participants in the experiment.} \label{fig1}
\end{figure}

\begin{figure}[tbp]
\centering
\includegraphics[width=0.6\textwidth]{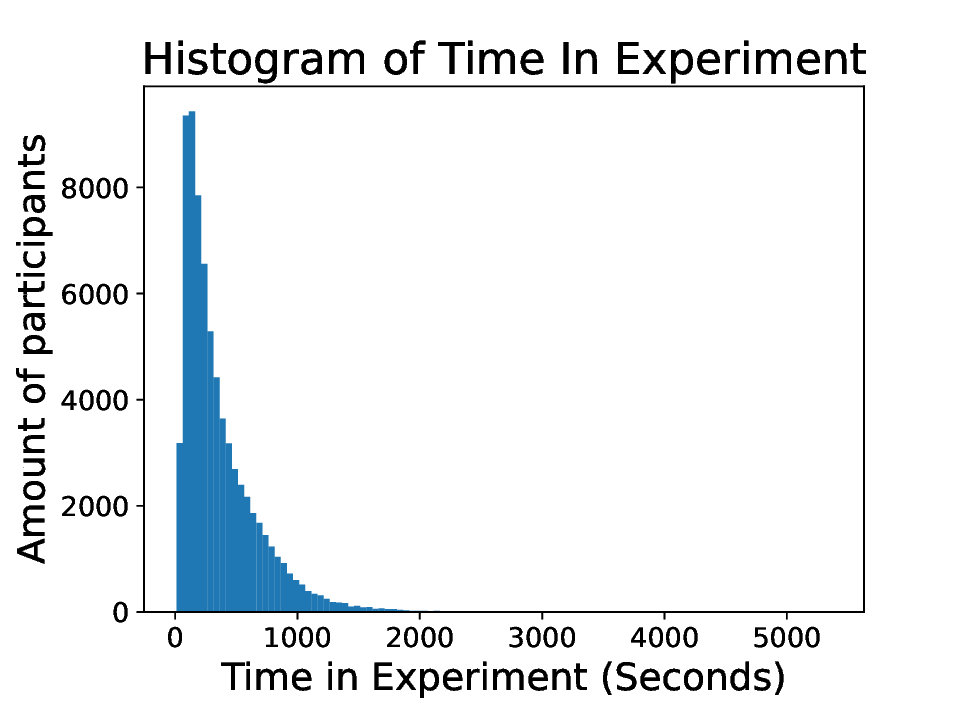}
\caption{Distribution of the time in the experiment.} \label{fig2}
\end{figure}

\begin{figure}[tbp]
\centering
\includegraphics[width=0.6\textwidth]{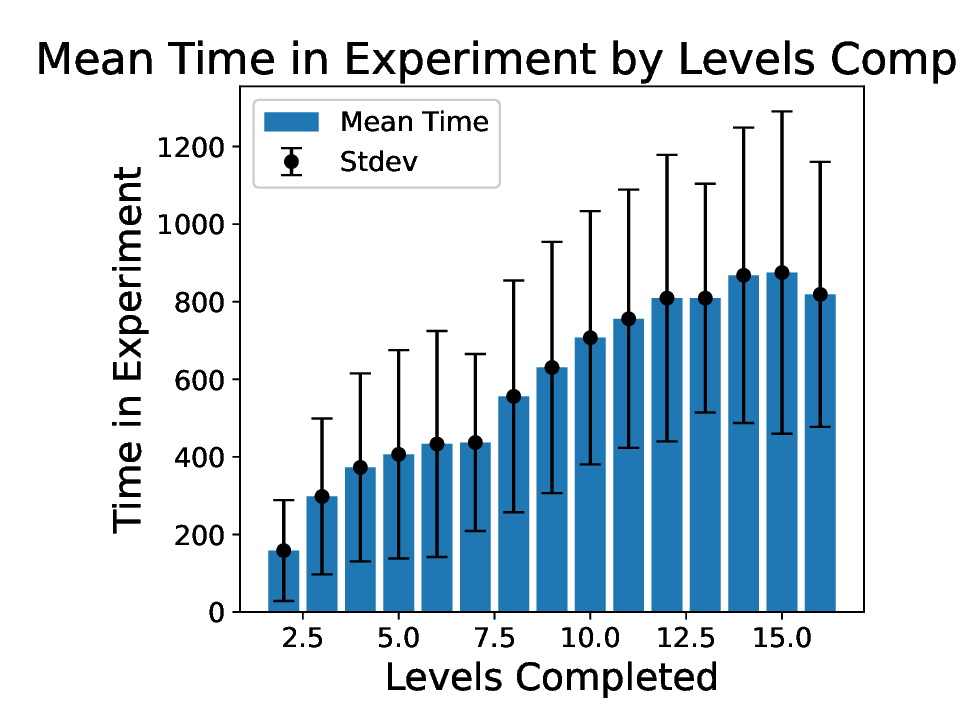}
\caption{Mean time in experiment grouped by levels completed.} \label{fig3}
\end{figure}

In Figure~\ref{fig1} the progress of participants over the experiment is displayed. In Figure~\ref{fig2}, the time spent in the experiment before either leaving or completing the experiment is displayed. The mean time spent in the experiment is 249.4 ($\sigma$ = 287.6), while the median time spent is 142 seconds. The mean time including the standard deviation is plotted in the error bar displayed in Figure~\ref{fig3}.

53,099 (42.14\%) participants left the experiment during the first level. This implies that participants did not finish a level. Since they did not finish a level, participants never completed a task. This data is left out of our analysis because there is no indication that participants did understand the task well enough or that their motivation was affected by the treatment they were subjected to.

Based on previous observations within the game, it is not possible to complete a stage within 5 seconds when you did not play the level before (based on loading times, time to search the button and the time to click on it). If a participant completed a stage within this 5-second timeframe, it is likely that the participant used some kind of cheating or did know the location of the button beforehand. Therefore, we left out the data of players from which the seconds spent per level (the average time spent per level) is below 5.0 seconds. This was the case for 11 (0.01\%) participants. This makes the total sample for analysis consist of 72,903 participants, of which 8488 (11.64\%) completed the experiment.

In Table~\ref{tab:tab2}, an analysis of the difference between participants who did and did not complete the experiment could be found. 11.6\% of the participants completed all levels of the experiment, with the average participant completing 3.67 levels (when excluding participants who did not complete a single level).

\begin{table}[tbp]
    \centering
    \caption{Analysis of variables between completed/not-completed experiments.}
    \begin{tabular}{*{3}{p{0.3\textwidth}}}
    \hline
         & Completed & Not completed\\
         \hline
        Number of entries & 8480 & 64,423 \\
        Levels completed & $\mu = 16,0, \sigma = 0,0$  & $\mu = 3,67, \sigma = 2,47$\\
        Seconds spent per level & $\mu = 51,18, \sigma = 21,35$ & $\mu = 86,53, \sigma = 62,65$\\
        \hline
    \end{tabular}
    
    \label{tab:tab2}
\end{table}

As mentioned in the methodology, there is no information available about the age of individual participants. However, if we assume the participants in the experiment follow the same demographics as the game, nearly 50 per cent of the participants are 13 or younger, with an additional 17 per cent being aged between 13 and 17. There are no other game-level demographics available.

Out of the participants, 23,079 (31.66\%) participants were in the control group, 24,894 (34.15\%) participants played together with an AI and did not know they were playing with an AI and 24.930 (34.20\%) participants played together with an AI and knew they playing with an AI. The distribution of both the self-chosen avatar and the assigned AI avatar of the groups who played with an AI are listed in Table~\ref{tab:tab3}. Over 50\% of the participants chose the first avatar as their avatar.

\begin{table}[tbp]
\caption{Distribution over avatar types}

\centering
\begin{tabular}{lllllll}
\hline
\begin{tabular}[c]{@{}l@{}}AI avatar →\\ ↓ Participant    avatar\end{tabular} & Avatar 1 & Avatar 2 & Avatar 3 & Avatar 4 & Avatar 5 &                 \\
\hline
Avatar 1                                                                             & 7502     & 7293     & 7552     & 7322     & 7388     & 37,057 (50.83\%) \\
Avatar 2                                                                             & 1078     & 1036     & 1114     & 1033     & 1025     & 5286 (7.25\%)    \\
Avatar 3                                                                             & 1762     & 1669     & 1634     & 1821     & 1726     & 8612 (11.81\%)   \\
Avatar 4                                                                             & 2596     & 2677     & 2699     & 2578     & 2622     & 13,172 (18.07\%) \\
Avatar 5                                                                             & 1770     & 1753     & 1711     & 1776     & 1766     & 8776 (12.04\%) \\
\hline
\end{tabular}
\label{tab:tab3}
\end{table}

To assess the gender of the avatars, we took the mean of all rankings given for that particular avatar. If the average avatar score was under 3, it was categorized feminine, while it was categorized masculine when the average avatar score was over 3. The categorization of all avatar types is listed in Table~\ref{tab:tab4}.

\begin{table}[tbp]
\centering
\caption{Categorization of avatars}
\begin{tabular}{lll}
\hline
Avatar & Average score & Categorization \\
\hline
1 & 1.903 ($\sigma = 1,26$) & Feminine \\
2 & 2.026 ($\sigma = 1,17$) & Feminine \\
3 & 3.509 ($\sigma = 1,27$) & Masculine \\
4 & 3.721 ($\sigma = 1,42$) & Masculine \\
5 & 4.051 ($\sigma = 1,41$) & Masculine\\
\hline
\end{tabular}
\label{tab:tab4}
\end{table}

\subsection{Hypothesis Tests}
\begin{figure}[tbp]
\centering
\includegraphics[width=0.6\textwidth]{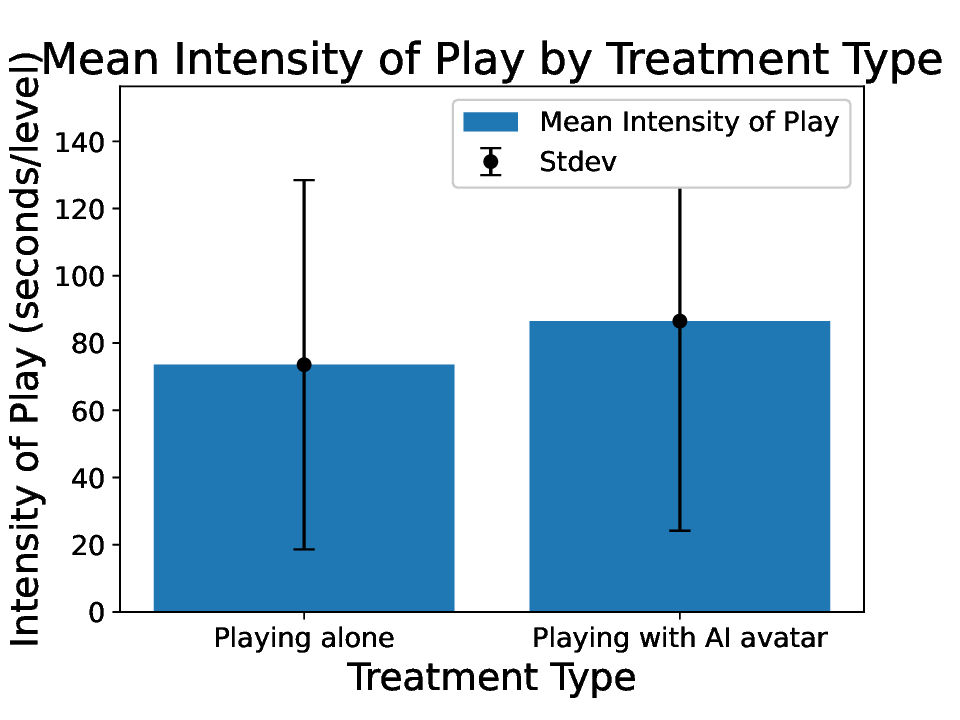}
\caption{Seconds spent per level grouped by treatment type.} \label{fig7}
\end{figure}

We conducted an ordinary least squares regression analysis to estimate the relationship between the presence of an AI avatar (treatment group) and the seconds spent per level (H1: Participants who play with an AI avatar (treatment group) have a higher intensity of playing than those who play alone (control group)). In Figure~\ref{fig7}, the seconds spent per level grouped by the treatment type is displayed.

We found a significant effect (R2 = 0.01, F(1, 72901) = 734.4, p < 0.001, 95\% CI [12.033, 13.909]) for intensity of play. Participants’ seconds spent per level increased by 13.0 seconds per level when playing together. That is equivalent to 0.22 standard deviations. This means that participants who played together with another (AI) avatar played less intensely. This is the inverse of what we expected, and therefore we reject H2b.

To test the influence of the effect of the gender appearance of the AI avatar, we conducted an ordinary least squares regression analysis (H2: Participants playing with an AI avatar that appears more masculine have a higher intensity of playing).

We found a significant effect for seconds spent per level (R2 = 0.000, F(1, 72901) = 10.03, p = 0.002, 95\% CI [0.841, 3.570]). Participants’ seconds spent per level increased by 2.2 seconds per level when playing together with a more masculine AI avatar. That is equivalent to 0.04 standard deviations. This means that participants who played together with an AI avatar that appeared more masculine played less intensely. This is the inverse of what we expected, and therefore we reject H2.

\begin{figure}[tbp]
\centering
\includegraphics[width=0.6\textwidth]{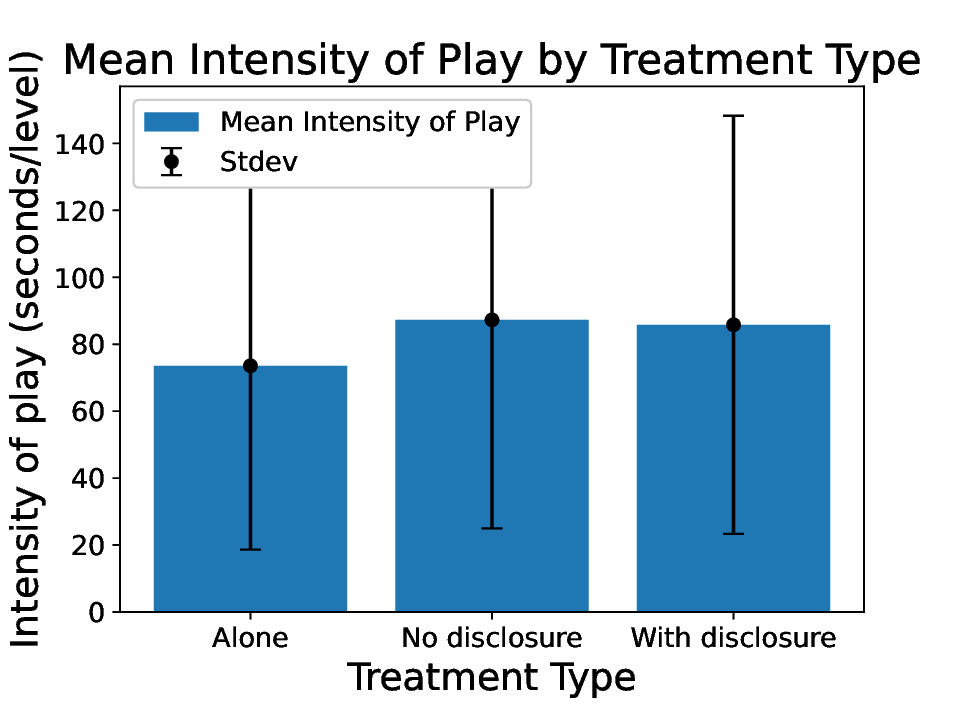}
\caption{Intensity of play by treatment type.} \label{fig10}
\end{figure}

To test the hypothesis about the effect of participants knowing that they are playing with an AI avatar on their intensity and duration of play, we conducted an ordinary least squares regression analysis (H3: Intensity of play is negatively associated with participants knowing that they are playing with an AI avatar). In Figure~\ref{fig10}, the seconds spent per level grouped by the treatment type is displayed.

We found a significant effect (R2 = 0.000, F(1, 49822) = 6.957, p = 0.008, 95\% CI [-2.569, -0.379]) for seconds spent per level. Participants’ seconds spent per level decreased by 1.5 seconds per level when the other avatar was disclosed as AI. That is equivalent to 0.02 standard deviations. This means that participants who played together with an AI avatar that was disclosed as such, play more intensely. This is opposing to what we expected.

In addition, we did not find a significant effect (R2 = 0.000, F(1, 48007) = 0.803, p = 0.370, 95\% CI [-0.003, 0.008]) that playing together with an AI avatar disclosed as such increases the value of seconds spent per level, compared to playing alone.

\section{Discussion}
This study researched various independent variables that could be used by an avatar controlled by Artificial Intelligence (AI) to improve users’ motivation. Our findings show that in our experiment and game, the masculinity of the users’ avatar makes participants play more intensely. Furthermore, playing together with another avatar controlled by AI significantly affects users’ motivation. Besides that, the masculinity of the AI’s avatar significantly increases the intensity of play. When users learn that the other avatar is controlled by AI, users had a higher probability of not completing the experiment and played more intensely. Finally, the difference between the user’s gender appearance and AI’s avatar gender appearance moderates the effect of the AI’s avatar gender appearance and makes users play more intensely.

\subsection{Contribution to existing literature}
Previous studies have shown that cooperating with others can increase motivation~\cite{featherstone2019unicraft, peng2013playing}. Our research shows that playing together with an avatar controlled by an AI makes players play more intense. Our findings suggest that the existing theory, which states that playing with someone else increases motivation, can be extended to include playing with an avatar controlled by AI. Future research in this field could focus on replicating our findings in other games and bringing these effects to non-gaming virtual worlds, to see if our findings are also applicable to other settings, such as virtual workspaces.

\subsection{Practical implications}
Companies that create virtual worlds, such as Meta, could use this research when designing avatars in the virtual world. If the sole goal would be to maximize player intensity, it would be best to use a masculine avatar for the AI and to tell users that they are interacting with an AI, in order to motivate users in their virtual world. However, the results of this study also show the need for diversity in virtual worlds. We have seen that the similarity between the participants’ avatar gender appearance and the AI avatar gender appearance moderates the effects of the AI avatar gender appearance on motivation. Therefore, preset genders of both the own avatar gender and the AI avatar gender can limit the effect of the AI avatar on motivation. Diversity in both avatars, such as being able to choose the avatars from a diverse list or being able to freely create the avatars, could help the user in motivation within the virtual world. Beyond gender, ensure that the avatars communicate attributes players are looking for, especially if you want to promote gender diversity in AI collaborators, whilst still stimulating motivation. This requires more research to investigate what other attributes might affect the motivation of players, as well as a broader coverage of the gender continuum, for instance by introducing more androgynous characters.

\subsection{Limitations}
Our AI avatar movement was simple. The avatar moved to preset points and did not interact with participants in another way, such as talking or following them. This could have limited the degree to which participants bestow agency to the AI. Furthermore, we don't theorize or study how the details of the interaction could change the user's perception regarding the agency of the AI ~\cite{vanneste2024artificial}. To address this limitation, future research could use a more naturally behaving AI, introduce different levels of intelligence, and observe the impact on the user's behaviour.

Besides the quality of the AI movement, the task and its setting might have influenced the outcomes of the study. The task was a simple search task, which could be seen as a more mathematical task, and the levels were themed around science. This particular task and setting could have led to a higher base motivation of some groups of participants, which could have worked through the measurements. Future research could be changing the task and setting, while keeping the rest of the methodology the same, to investigate if the task and setting influences the effects we found.

Finally, the selection of avatars was limited to only five avatars, which restricted the gender expressions in the avatars and diversity in skin colour. A participant raised this concern, which did not feel represented by the available avatars due to skin color. We decided to not include diverse avatars because participants should choose based on gender, not on other avatar features. However, this non-diverse set of avatars could have led to dissatisfaction with the experiment and decreased motivation regardless of the treatment type for some participants. To address this limitation, future research could include a wider range of avatars to choose from or allow players to use their avatar.

\section{Conclusion}
This study aimed to describe how AI avatars inside virtual worlds could be used to increase the motivation of a user. We conducted a large-scale field study inside a game where we were able to control variables in the virtual world. We found that playing together with an AI avatar makes users play less intensely. When this AI avatar appears more masculine, users play less intensely. The AI avatar being disclosed as an AI make users play more intensely. Further research could include replicating these results in other games \& non-gaming virtual worlds and researching what the causes are for the effects we found.

\begin{credits}

\subsubsection{\discintname}
Boele Visser is co-owner of the game in which the research took place. The other authors have no competing interests.
\end{credits}

%\newpage
\bibliographystyle{splncs04}
\bibliography{list}
%

% \AtNextBibliography{\small}
% \printbibliography[]

\end{document}